# Probabilistic Diffusion Models Advance Extreme Flood Forecasting

**Zhigang Ou[1,2]* Congyi Nai[2]*, Baoxiang Pan[2]†, Yi Zheng[1,3]†, Chaopeng Shen[4], Peishi Jiang[5], Xingcai Liu[6], Qiuhong Tang[6], Wenqing Li[7], and Ming Pan[8]**

[1]School of Environmental Science and Engineering, Southern University of Science and Technology, Shenzhen, China.

[2]Key Laboratory of Earth System Numerical Modeling and Application, Institute of Atmospheric Physics, Chinese Academy of Science, Beijing, China.

[3]State Environmental Protection Key Laboratory of Integrated Surface Water-Groundwater Contamination Control, School of Environmental Science and Engineering, Southern University of Science and Technology, Shenzhen, China.

[4]Civil and Environmental Engineering, The Pennsylvania State University, University Park, PA, USA.

[5]Atmospheric, Climate, and Earth Sciences Division, Pacific Northwest National Laboratory, Richland, WA, USA.

[6]Key Laboratory of Water Cycle and Related Land Surface Processes, Institute of Geographic Sciences and Natural Resources Research, Chinese Academy of Sciences, Beijing, China.

[7]State Key Laboratory of Stimulation and Regulation of Water Cycles in River Basins, China Institute of Water Resources and Hydropower Research, Beijing, China.

[8]Scripps Institution of Oceanography, University of California San Diego, La Jolla, CA, USA.

*****Equal contribution.

**†**Corresponding authors:

Yi Zheng (zhengy@sustech.edu.cn);

Baoxiang Pan(panbaoxiang@lasg.iap.ac.cn)

**Key Points:**

- DRUM, a diffusion-based generative AI model, enables reliable probabilistic forecasting of extreme floods without predefined distributions
- DRUM outperforms LSTM-based methods and extends early warnings for extreme floods across CONUS under real-world conditions
- Reliable weather forecasts are vital for flood prediction, with their value in improving early warnings varying by flood mechanisms

## Abstract

Extreme floods pose escalating risks in a changing climate, yet forecasting remains challenging due to peak flow underestimation and high uncertainty. We introduce DRUM, a diffusion-based probabilistic deep learning approach that advances extreme flood forecasting across representative basins in the contiguous United States. DRUM outperforms state-of-the-art benchmarks, enhancing nowcasting skill for the top 1‰ of flows in 72.3% of studied basins. Under operational scenarios, DRUM extends reliable lead times by nearly a full day for 20- and 50-year floods. When evaluated with measured precipitation, an ideal condition, recall improves by 0.3–0.4 and the early warning window extends by 2.3 days for 50-year floods. The enhancement potential varies regionally, with precipitation-driven flood zones in the eastern and northwestern U.S. benefiting most, gaining 3–7 days in lead time. These findings highlight the transformative potential of diffusion models as a cutting-edge generative AI technique for advancing hydrology and broader Earth system sciences.

## Plain Language Summary

Extreme floods are becoming more frequent and destructive due to climate change, yet predicting them remains difficult. Existing models often underestimate peak river flows and fail to effectively quantify uncertainty due to the scarcity of peak flow samples in historical datasets. In this study, we introduce DRUM, a new deep learning approach based on diffusion models, a cutting-edge generative AI technique. Tested in representative U.S. river basins, DRUM outperforms state-of-the-art deep learning models for the most extreme floods in over 70% of cases and extends early warnings by nearly a full day for major floods. When given actual rainfall data instead of weather forecasts, DRUM's accuracy further improves, increasing recall by 0.3–0.4 and extending early warnings by 2.3 days for the most extreme floods. In regions where floods are mainly driven by heavy rainfall, such as the eastern and northwestern United States, it further extends warning times by up to 7 days. These results demonstrate how AI can enhance flood prediction and disaster preparedness while highlighting the importance of accurate rainfall forecasts. Beyond flood forecasting, this study illustrates the broader potential of generative AI in improving predictions of extreme environmental events and advancing Earth system science.

## 1 Introduction

Since 2000, catastrophic flood events have increased by 134%, causing over 100,000 deaths and USD 651 billion in economic damage (UNDRR, 2020). This escalating impact is driven by both the rising frequency of floods (Yin et al., 2018) and continued urbanization in flood-prone areas (Rentschler et al., 2023). Reliable river flood forecasting remains elusive, primarily due to persistent underestimation of peak flows and inadequate uncertainty quantification (Auer et al., 2024; Bárdossy & Anwar, 2022; Jain et al., 2018), which undermines risk warnings, often leading to misinformed decisions and ineffective emergency responses. While recent advances in observations, modeling, and computation have improved flood forecasting (Cui et al., 2024; Frame et al., 2022; Nearing et al., 2024), critical gaps remain, highlighting the urgent need for more reliable and actionable river flood forecasts.

Leveraging extensive historical data, Deep Learning (DL), particularly Long Short-Term Memory (LSTM) networks (Feng et al., 2020; Kratzert et al., 2018, 2019; Nai et al., 2024), has achieved unprecedented predictive accuracy in river flow prediction. However, the scarcity of extreme flood samples in historical datasets may lead deterministic DL models to underestimate peak flows (Brunner et al., 2022; Xie et al., 2021), emphasizing the need for probabilistic forecasting to address uncertainty. Traditional uncertainty quantification methods, such as parametric distribution fitting, Monte Carlo simulation, and approximate Bayesian computation, struggle with rigid assumptions and convergence issues (Darbandsari & Coulibaly, 2021; Sadegh & Vrugt, 2013; Shi et al., 2023), limiting their applicability to modern DL frameworks. While probabilistic LSTM (Klotz et al., 2022) advances uncertainty quantification by directly generating runoff distribution parameters through end-to-end uncertainty learning, its reliance on prespecified parametric assumptions restricts flexibility. Probabilistic diffusion models (Ho et al., 2020; Ho & Salimans, 2022), a cutting-edge generative AI technique, iteratively denoise random signals through theoretically-grounded probabilistic frameworks, eliminating dependency on predefined distributions. While diffusion models have demonstrated breakthroughs in meteorological forecasting (Li et al., 2024; Price et al., 2025), their transformative potential for broader geoscientific applications, including runoff prediction, remains unexplored.

This study develops the diffusion-based runoff model (DRUM), which leverages meteorological forcings and static catchment attributes for probabilistic runoff predictions. By evaluating DRUM's performance against state-of-the-art LSTM-based models across the contiguous United States (CONUS), we aim to address three key questions: (1) Can generative AI enhance the prediction of extreme floods compared to existing deep learning methods? (2) How does DRUM perform under real-world operational conditions where meteorological forecasts introduce uncertainty? (3) To what extent does improved precipitation input strengthen extreme flood early warning capability? Our results provide new insights into the role of generative AI in hydrology and highlight its potential for advancing flood risk management in a changing climate.

## 2 Materials and Methods

### 2.1 DRUM

DRUM is a conditional diffusion model that generates *N*-day probabilistic runoff forecasts $X$ by incorporating $t$-day historical dynamic forcings $H$, *N*-day precipitation forecasts $F$, and static catchment attributes $S$. The diffusion framework establishes a bijective mapping between simple Gaussian distributions and complex streamflow distributions through a forward–backward process (Figure 1). The forward process progressively degrades a streamflow sample $x_0$ from the data distribution $q(x_0)$ into a Gaussian noise $x_T$ from the Gaussian distribution $q(x_T)$ by adding Gaussian noise $\varepsilon$, which can be defined as:

$$q(x_t|x_{t-1}) = N\left(x_t; \sqrt{(1-\beta_t)}x_{t-1}, \beta_t \boldsymbol{I}\right), \tag{1}$$

where $\beta_t$ is a predefined variance schedule controlling the noise level at each noise level step $t$, with $t$ ranging from 1 to $T(T = 1200)$.

In the reverse process, a noise prediction model is trained to approximate the conditional probability $p_\theta(x_{t-1}|x_t, H, S, F)$, reconstructing the streamflow distribution iteratively. Here, $p_\theta$ represents the likelihood of the previous state $x_{t-1}$ given the current state $x_t$ and conditional inputs $y = (H, S, F)$. This can be formulated as:

$$p_\theta(x_{t-1}|x_t, y) = N\big(x_{t-1};\ \mu_\theta(x_t, t, y), \sigma_\theta^2(x_t, t, y)\boldsymbol{I}\big), \tag{2}$$

where $\mu_\theta$ and $\sigma_\theta^2$ are learned functions parameterized by $\theta$.

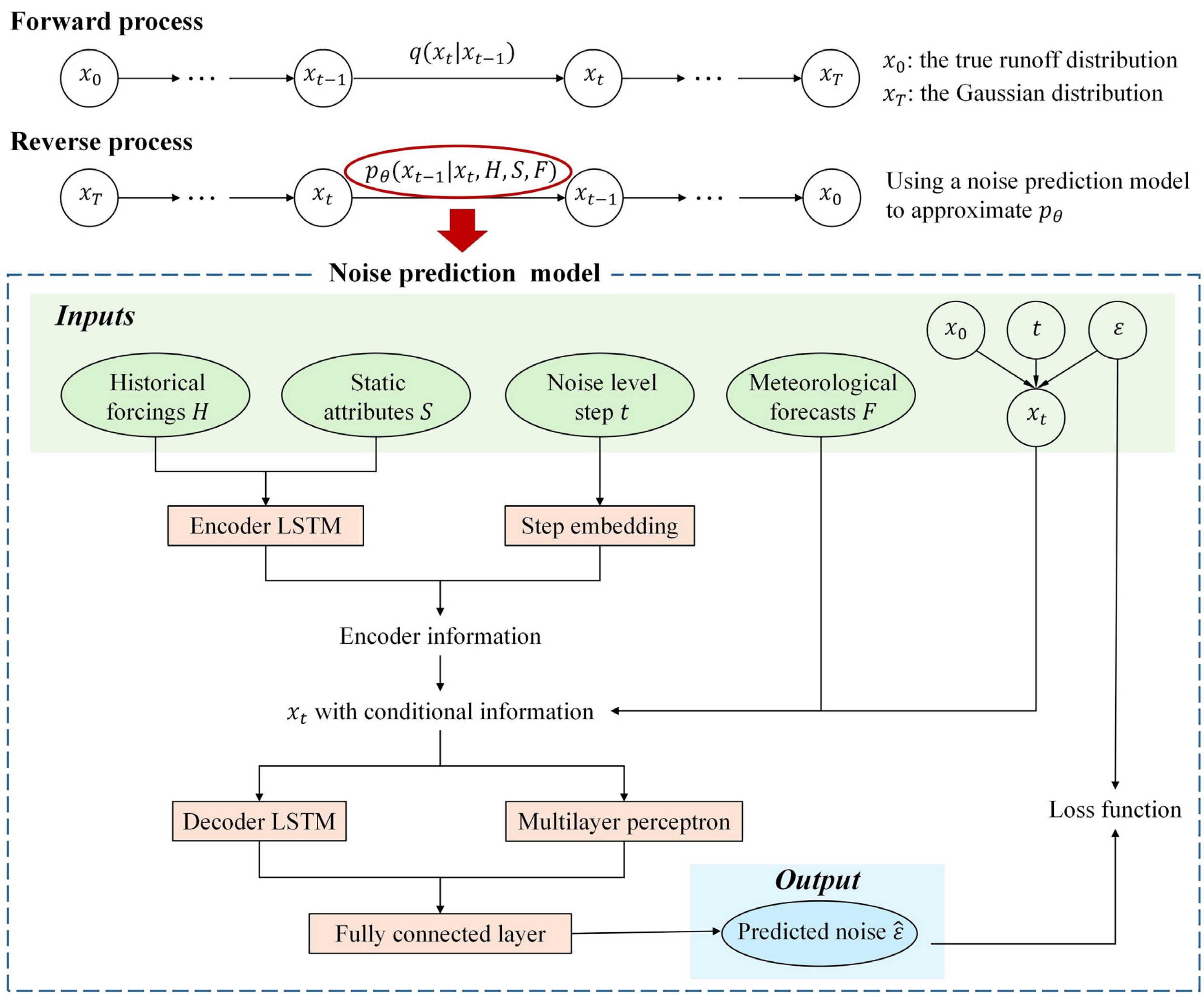

**Figure 1. DRUM's forward–backward process and the structure of its noise prediction model.** The forward process follows a Markov chain that progressively injects Gaussian noise $\varepsilon$ into streamflow, while the reverse process reconstructs distribution through iterative denoising. The noise prediction model's backbone is an encoder–decoder LSTM, supported by an embedding layer and a multilayer perceptron. The model inference and sampling procedure are shown in Figure S1 in the Supporting Information (SI).

Diffusion models learn to estimate the noise $\hat{\varepsilon}_t$ at noise level step $t$, which is equivalent to learning the score function $\nabla_{x_t} \log p(x_t)$ (Song & Ermon, 2019). Based on Bayes' rule:

$$\nabla_{x_t} \log p\,(x_t|y) = \nabla_{x_t} \log p\,(y|x_t) + \nabla_{x_t} \log p\,(x_t), \tag{3}$$

To facilitate conditional sampling, this score function can be reformulated by introducing a guidance weight $\gamma$ (Dhariwal & Nichol, 2021):

$$\nabla_{x_t} \log p\,(x_t|y) = \gamma \nabla_{x_t} \log p\,(y|x_t) + \nabla_{x_t} \log p\,(x_t), \tag{4}$$

where $\nabla_{x_t} \log p\,(x_t)$ is an unconditional diffusion model and $\nabla_{x_t} \log p\,(y|x_t)$ is a classifier. Using Bayes' rule, $p(y|x_t)$ can be expressed as:

$$p(y|x_t) = \frac{p(x_t|y) \cdot p(y)}{p(x_t)}, \tag{5}$$

$$\nabla_{x_t} \log p\,(y|x_t) = \nabla_{x_t} \log p\,(x_t|y) - \nabla_{x_t} \log p\,(x_t). \tag{6}$$

Substituting Equation (6) into Equation (4) and introducing a modified guidance weight $\gamma + 1$:

$$\nabla_{x_t} \log p_\gamma\,(x_t|y) = (1+\gamma)\nabla_{x_t} \log p\,(x_t|y) - \gamma \nabla_{x_t} \log p\,(x_t). \tag{7}$$

Equation (7) highlights how $\gamma$ adjusts the strength of the conditional signal. Specifically, $\gamma = -1$ yields unconditional sampling, $\gamma = 0$ corresponds to standard conditional formulation, and $\gamma > 0$ amplifies conditional signal's influence on generated samples. This approach bridges unconditional and conditional diffusion models, enabling flexible control over the balance between prior and conditional information.

Detailed derivations and the inference procedure of DRUM are provided in Text S1, and Text S2 further explains how diffusion models mitigate data imbalance.

## 2.2 Datasets

The Catchment Attributes and MEteorology for Large-sample Studies (CAMELS) dataset (Addor et al., 2017; Newman et al., 2015) includes 671 basins across CONUS, representing diverse geological, ecological, and climatic conditions. It provides daily meteorological forcings (1980–2010) from three gridded products—Daymet (0.01°), Maurer (0.125°), and NLDAS (0.125°) —along with 27 static catchment attributes (Table S1) and daily streamflow observations at basin outlets. Following Klotz et al. (2022), we use the 531 basins selected by Newman et al. (2017), who excluded basins larger than 2000 km² and those with large area discrepancies (>10%) between different USGS geospatial datasets.

The European Centre for Medium-Range Weather Forecasts Integrated Forecasting System (ECMWF-IFS) reforecasts dataset (Vitart et al., 2017) provides precipitation forecasts at 0.125° resolution, generated twice weekly within a 20-year rolling window. Each forecast consists of four perturbed ensemble members with 6-hourly accumulations extending up to 7 days ahead. We use reforecasts from 30 September 1995 to 1 October 2005, aggregating 6-hourly data into daily totals. Missing values (~10%) are filled using ERA5-Land reanalysis dataset (Muñoz Sabater, 2019). All gridded data are area-weighted to obtain basin-scale estimates.

**2.3 Numerical experiments**

Previous studies on the CAMELS dataset have shown that well-tuned LSTM models (median NSE: 0.73–0.77) generally outperform traditional process-based models (median NSE: 0.55–0.68) in predictive accuracy for rainfall-runoff modeling (Frame et al., 2023; Kratzert et al., 2019). Against this established context, we compare DRUM with two LSTM-based baseline models: a deterministic model (LSTM-d) and a probabilistic model (LSTM-p). The three models share a similar backbone architecture, comprising an encoder-decoder LSTM with static catchment attribute embedding components (Table S2). The models take as input 365-day historical meteorological forcing sequences, 27 static catchment attributes, and *N*-day precipitation forecasts to generate *N*-day streamflow predictions. DRUM leverages a diffusion process for probabilistic forecasting, LSTM-d produces deterministic forecasts optimized with a basin-averaged Nash–Sutcliffe efficiency loss function (Kratzert et al., 2019), while LSTM-p characterizes uncertainty through a single asymmetric Laplace distribution (Nearing et al., 2024) by outputting parameters at each forecast step.

We design two experiments for model comparison, focusing on nowcasting and operational forecasting, respectively. In **Experiment 1**, we evaluate single-step streamflow nowcasting (0-day lead time, i.e., forecasting the present day) using merged meteorological forcings from Daymet, Maurer, and NLDAS, which showed optimal performance (Kratzert et al., 2021). Input features include precipitation, maximum temperature, minimum temperature, shortwave radiation, and vapor pressure. Following Klotz et al. (2022), we split the dataset into training (10/01/1980–09/30/1990), validation (10/01/1990–09/30/1995), and testing periods (10/01/1995–

09/30/2005). **Experiment 2** extends to operational forecasting using future precipitation forecasts. We generate 7-dimensional output sequences (1–7-day lead times) with Daymet data for the same meteorological features as Experiment 1. During training, Daymet precipitation observations are treated as ground truth (i.e., measured precipitation), representing the best available observational forcing data. For testing, we evaluate model performance using both measured precipitation and ECMWF-IFS reforecasts while maintaining the same data split as Experiment 1. The primary objective of this design is to quantitatively assess how improvements in meteorological forecast accuracy translate into gains in flood forecasting performance.

The probabilistic models (DRUM and LSTM-p) generate 50 ensemble forecasts per time step to characterize prediction uncertainty, whereas the deterministic model (LSTM-d) produces single-valued outputs. Model evaluation compares predictions against observations, using all ensemble members for probabilistic metrics and ensemble means for deterministic metrics.

DRUM, LSTM-p, and LSTM-d took 16.5, 2.5, and 4.8 hours to train, respectively, on an NVIDIA RTX 4090 GPU. For predictions, DRUM required 10 minutes per basin to generate 50 samples over the 10-year test horizon, while LSTM variants had negligible sampling time. It worth noting that DRUM's efficiency can be improved 20–50× using Denoising Diffusion Implicit Models (Song et al., 2020) and further accelerated with multi-GPU parallelization.

### 2.4 Evaluation metrics

Continuous Ranked Probability Score (CRPS) is a metric for assessing probabilistic forecasts (Gneiting & Raftery, 2007), measuring both calibration and sharpness by evaluating the entire forecast distribution rather than single-point predictions. For empirical distributions of finite support, the discrete form of CRPS is:

$$\text{CRPS} = \sum_{i=1}^{n} P(y_i)^2 \cdot I(y_i \le x_{\text{obs}}) + \sum_{i=1}^{n} \left(P(y_i) - 1\right)^2 \cdot I(y_i > x_{\text{obs}}), \tag{8}$$

where $P(y_i)$ is the CDF of the discrete forecast value $y_i$, $n$ is the number of forecast values, and $I(\cdot)$ is an indicator function. To evaluate forecast performance over an extended period, the average CRPS is calculated as:

$$\text{CRPS}_{\text{avg}} = \frac{1}{T}\sum_{t=1}^{T} \text{CRPS}_t\,, \tag{9}$$

where $T$ is the number of time points in the evaluation period. CRPS reduces to Mean Absolute Error (MAE) for single-member forecasts and ranges from 0 (perfect forecast) to infinity.

We assess flood detection skill using precision, recall, and F1 score (range: 0–1; 1 = perfect performance). Flood magnitudes are defined by return periods following the USGS Bulletin 17B, with a Log-Pearson Type III distribution fitted to annual maximum daily streamflow series for each basin. A flood event is correctly forecast if observed and predicted peak flows occur on the same day and exceed the return period threshold. Precision quantifies the proportion of true floods among all predicted events, while recall measures the fraction of actual flood events successfully identified. The F1 score, the harmonic mean of precision and recall, provides a balanced assessment.

We also measure early warning capability using mean lead time (in days). For each flood event, the maximum lead time at which the forecast detects the event is recorded; if the event is undetected, it is assigned a lead time of zero. Basin-specific warning capability is represented by the mean lead time across all flood events. The early warning capacity for floods of different magnitudes (e.g., return periods) can be similarly evaluated using mean lead time.
Details of all evaluation metrics can be found in Text S3.

## 3 Results

### 3.1 Superior nowcasting performance for extreme floods

We benchmark DRUM against deterministic (LSTM-d) and probabilistic (LSTM-p) LSTMs across 531 CONUS basins using the CAMELS dataset. DRUM outperforms these state-of-the-art models in nowcasting in both deterministic (Table S3) and probabilistic (Figure 2) perspectives. The improvement, measured by the CRPS, is particularly notable in probabilistic nowcasting, with gains increasing as flow magnitude rises (Figure 2a). For the top 1‰ of flows, DRUM surpasses LSTM-p in 72.3% of basins (Figure 2b) and LSTM-d in 76.3% (Figure S2), with all comparisons statistically significant (paired Wilcoxon tests, $p$<0.01). Spatial transferability analysis shows

DRUM and LSTM-p perform comparably across metrics (e.g., median KGE: 0.691 vs. 0.705, $p$=0.362), with DRUM superior under high-flow conditions (Text S4, Tables S4–S5 and Figure S3). DRUM's advantage is further evident in eight extreme flood events exceeding the range of training data (Figures 2c–j), where its ensemble mean closely aligns with observed peaks, and its 95% prediction intervals (PIs) effectively capture observed trajectories. In contrast, LSTM models severely underestimate flow peaks, with LSTM-p producing overly wide 95% PIs, limiting its probabilistic utility.

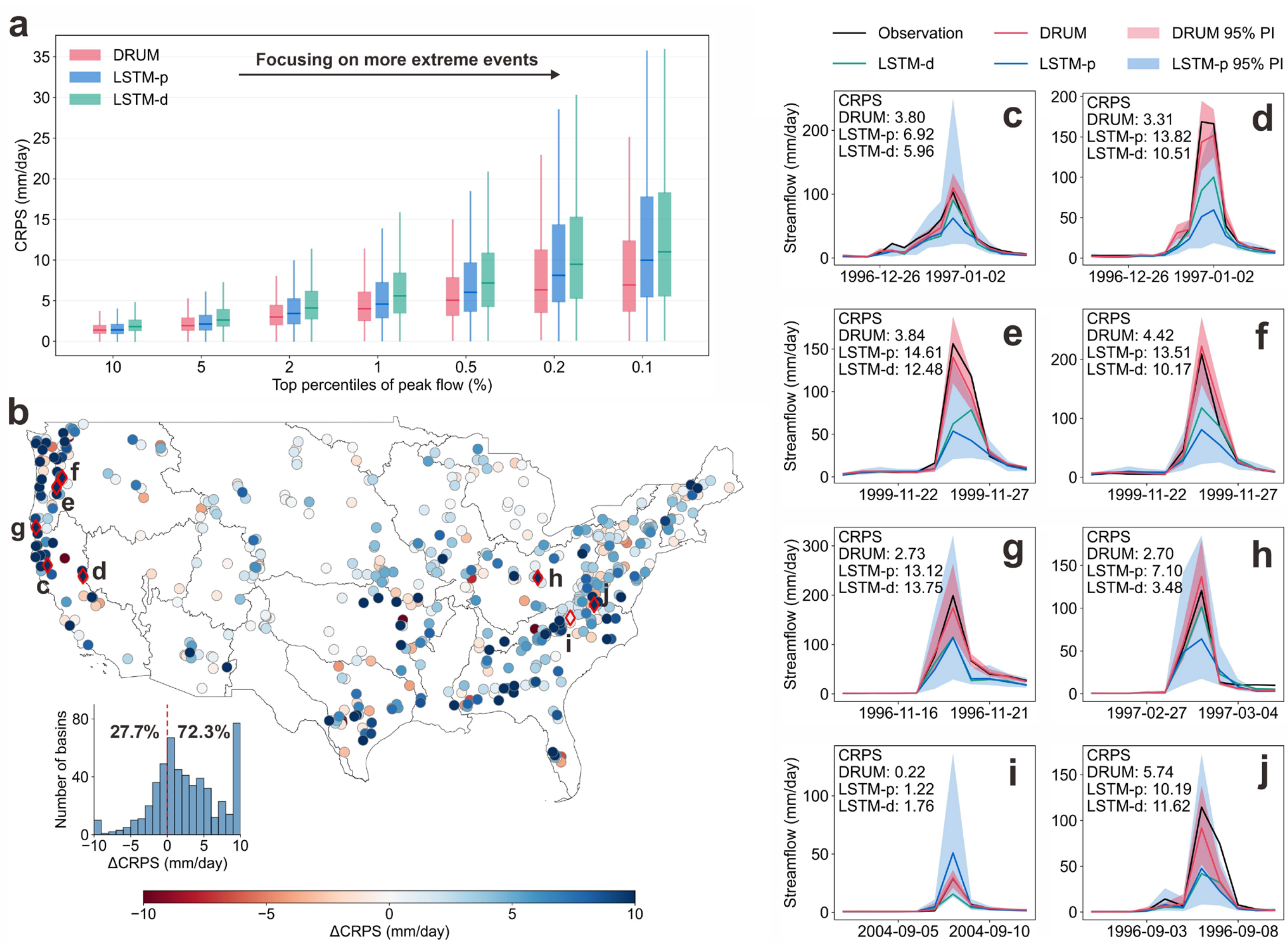

**Figure 2. Improved nowcasting of extreme flood events across 531 representative basins in the contiguous United States (CONUS). (a)** Comparison of Continuous Ranked Probability Score (CRPS) across top percentiles of peak flow among DRUM, LSTM-p, and LSTM-d. The horizontal line in each box indicates the median, with all values included in boxplot statistics, though outliers are excluded for clarity. **(b)** Comparison of CRPS between DRUM and LSTM-p for the top 1‰ of flows, where positive ΔCRPS (LSTM-p minus DRUM) indicates better performance by DRUM. Inset

shows the ΔCRPS distribution truncated to [−10, 10] for clarity. **(c–j)** Forecasts of top-ranked flow peaks in locations marked in Figure 2b. Solid lines represent ensemble means for probabilistic models and deterministic model predictions, with shaded areas indicating their 95% prediction intervals (PIs).

While DRUM and LSTM-p exhibit similar spatial patterns in uncertainty estimates (95% PIs; Figure 3a, Pearson's r > 0.8 in ~80% of basins), a spatial analysis using a scaling factor $k$ (DRUM = $k$ × LSTM-p) reveals that DRUM produces more concentrated uncertainty estimates (Figure 3b). This difference is most pronounced in extreme floods exceeding the training data range (Figure 3c–j), where DRUM generates sharper probability distributions centered around observed values, whereas LSTM-p yields notably wider distributions with higher CRPS. These findings highlight DRUM's ability to enhance both nowcasting accuracy and uncertainty quantification across all flow conditions, making a significant advancement in river flood forecasting.

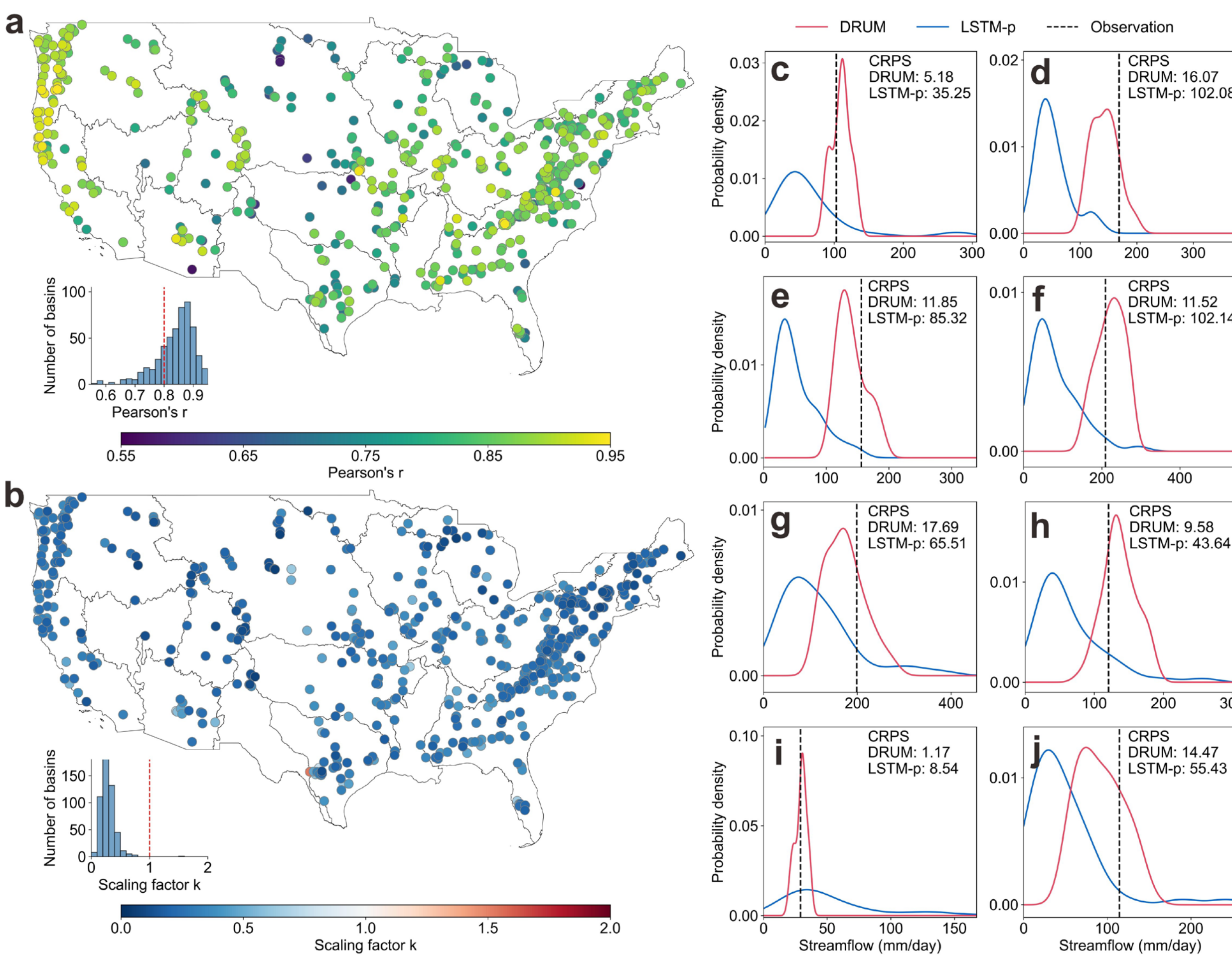

**Figure 3. Probabilistic models for uncertainty quantification in rainfall-runoff prediction. (a)** Spatial distribution of correlation (Pearson's $r$) of uncertainty estimates between DRUM and LSTM-p. All basins showing statistical significance ($p < 0.01$, two-tailed Student's t-test). **(b)** Spatial distribution of scaling factor $k$, where $k < 1$ indicates more concentrated uncertainty estimates from DRUM ($p < 0.01$). **(c–j)** Probability distributions of peak discharge forecasts during extreme floods. Basin locations and flood events correspond to Figure 2b and 2c–j, respectively.

**3.2 Robust operational forecasting under data uncertainty**

We generate 7-day streamflow predictions with DRUM and LSTM-based models, driven by ECMWF-IFS reforecasts, to simulate real-world operational conditions where meteorological inputs are uncertain. DRUM consistently outperforms LSTM-based models across flood magnitudes and lead times, as indicated by F1 scores (Figure 4a). It also demonstrates superior flood early warning capability (Figure 4b), particularly for extreme events (20- and 50-year floods), where it extends the mean lead time from ~0.2 to ~1.2 days—nearly a full day improvement. Two representative cases further illustrate DRUM's advantage under inaccurate precipitation forecasts. For a 5-year flood (Figure 4c), DRUM demonstrates higher accuracy with reduced uncertainty. For a 50-year flood (Figure 4d), it provides an effective 1-day warning, whereas LSTM-based models fail across all lead times.

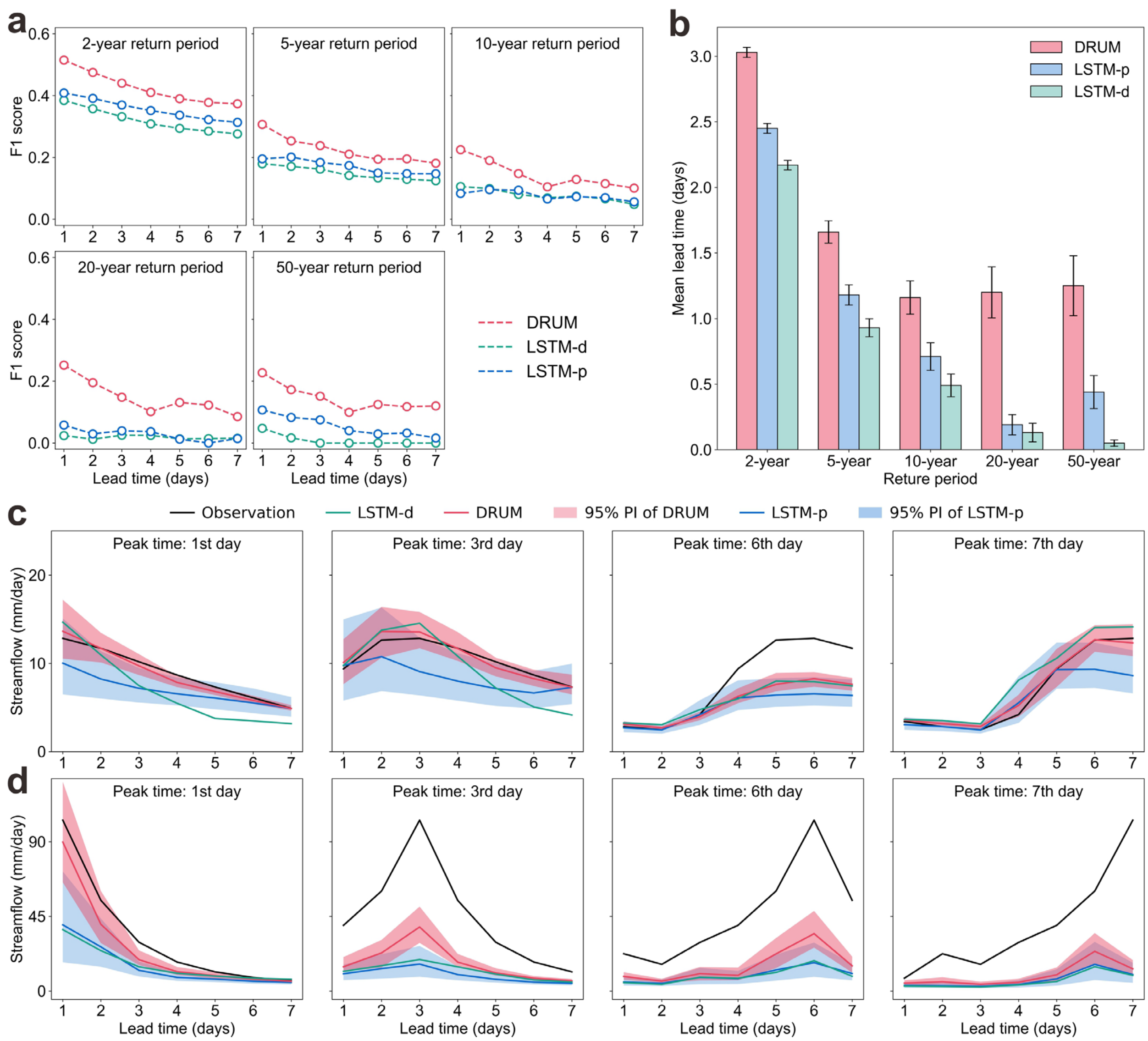

**Figure 4. Performance evaluation and case studies in operational flood forecasting. (a)** F1 score evolution with lead time for flood events of different magnitudes (2- to 50-year return periods). **(b)** Comparison of models of early flood warning capability, measured by mean lead time across return periods. Error bars represent mean ± standard error. **(c, d)** Forecast comparisons for a 5-year flood (c) and a 50-year flood (d) at selected locations (Figure S4), showing predictions 1–7 days before the flood peak.

### 3.3 Improved forecasting with accurate precipitation inputs

We further evaluated the impact of replacing ECMWF-IFS reforecasts with measured precipitation on flood forecasting. Under this ideal data condition, DRUM exhibits improvements in both recall and precision (Figure 5a). Recall increases with flood severity, with Δrecall reaching 0.3–0.4 for 50-year floods, highlighting greater potential for improvement due to systematic precipitation underestimation (Figure S5). This sensitivity in recall is particularly relevant for operational flood forecasting, where missing a flood event poses a greater risk than issuing a false alarm. Figure 5b exhibits the enhancement of mean lead time across different flood magnitudes, with the most severe 50-year flood exhibiting a 2.3-day extension in the early warning window. The benefit of improved precipitation data condition, measured by the difference in mean lead time, varies spatially (Figure 5c). The eastern and northwestern United States show the greatest enhancement potential, with improved precipitation accuracy extending warning times by 3–7 days. This pattern aligns with the dominant precipitation-driven flood mechanisms documented in these regions (Jiang et al., 2022; Stein et al., 2021). This finding underscores the value of ongoing efforts to advance meteorological prediction using deep learning (Bi et al., 2023; Price et al., 2024).

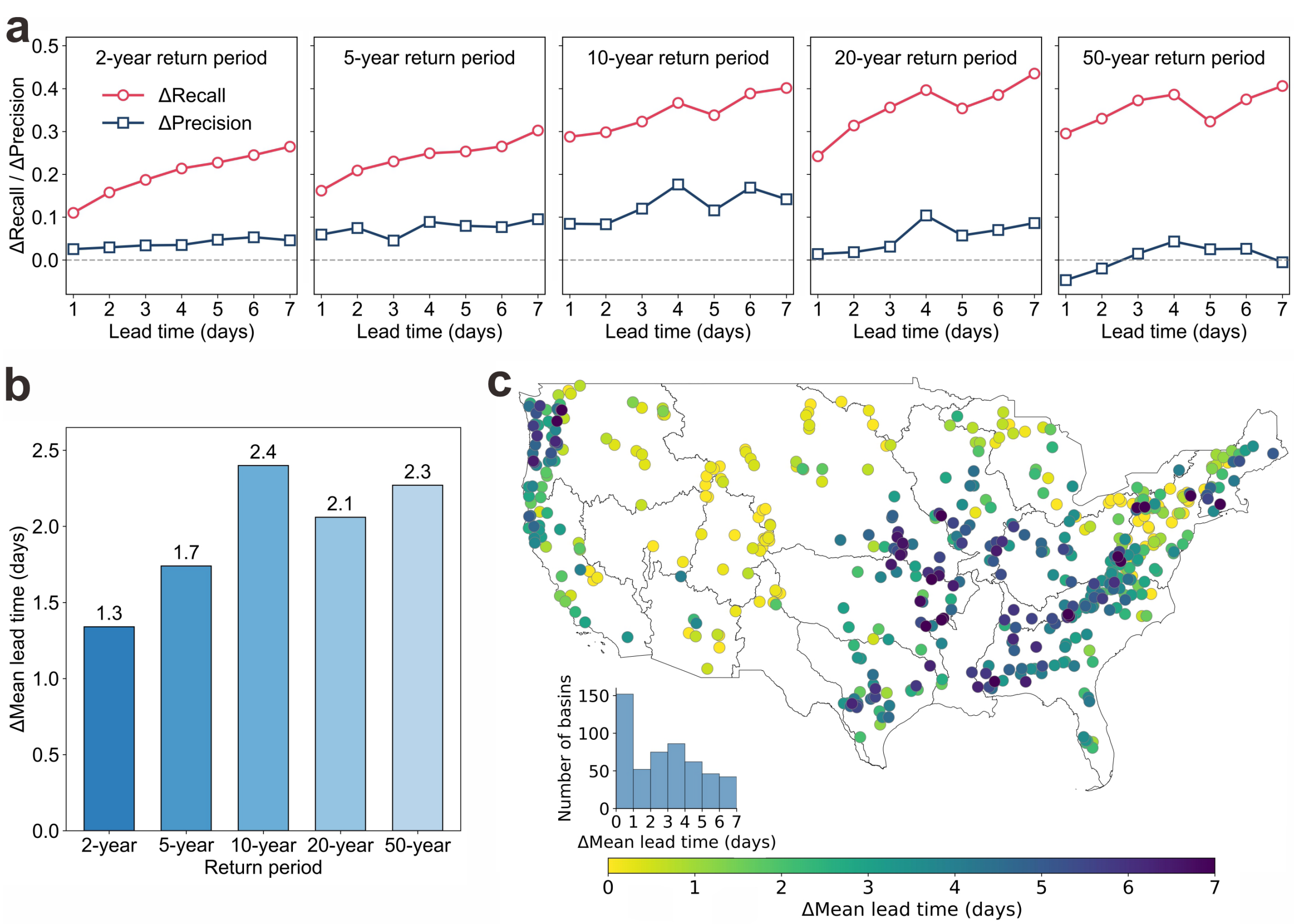

**Figure 5. Impact of precipitation forecast quality on flood forecasting skill and early warning capability. (a)** Improvement in recall and precision when using forecasts driven by measured precipitation, compared to those driven by ECMWF-IFS reforecasts, across different flood return periods (2-, 5-, 10-, 20-, and 50-year). **(b)** Enhancement in flood early warning capability, measured as increase in mean lead time across return periods. **(c)** Spatial distribution of flood early warning improvements. The inset histogram summarizes the distribution of ΔMean lead time across basins.

## 4 Discussion

DRUM's superior performance stems from three key aspects: distribution-free probabilistic modeling, multi-scale pattern decomposition, and flexible conditional generation. Together, these elements address fundamental limitations of traditional approaches, which rely on explicit probability distributions (e.g., Gaussian or asymmetric Laplace), often fail to capture the complex and often heavy-tailed nature of flood distributions. DRUM, through its diffusion framework, implicitly learns true data distributions through iterative denoising, eliminating rigid assumptions and excelling in capturing distributional tails (Figure S6). Such flexibility is crucial for river flood forecasting, where tail risks are often underestimated. DRUM's iterative refinement process decomposes flood forecasting into numerous learnable subtasks. The forward process adds noise progressively, while the reverse process reconstructs hydrological patterns iteratively. This hierarchical approach allows DRUM to effectively model diverse hydroclimatic dynamics. In contrast, LSTM models must handle all dynamics simultaneously in a single prediction step. Unlike LSTM-based models' fixed weights, DRUM uniquely balances conditional and unconditional generation through a weighting mechanism—classifier-free guidance (Ho & Salimans, 2022)—with the influence of the guidance weight on generation results (Table S6). This approach captures general empirical streamflow distributions while adapting to meteorological and catchment-specific conditions, maintaining prediction accuracy and physical consistency.

Comprehensive evaluations across various flood magnitudes and lead times (Figure 4) demonstrate that the diffusion process more effectively preserves and utilizes precipitation signals throughout generation. This advantage enables DRUM to consistently outperform LSTM-based models in flood forecasting, particularly for extreme events. The success of precipitation conditioning underscores the potential to incorporate additional hydrological signals. While precipitation is the primary driver of floods, flood generation is modulated by multiple interacting factors (Bertola et al., 2021; Jiang et al., 2024), such as temperature-driven snowmelt processes in cold regions (Blöschl et al., 2019) and soil moisture-regulated rainfall-runoff dynamics (Wasko & Nathan, 2019). Given DRUM's proven ability to leverage conditioning signals, diffusion models hold significant potential for integrating multiple inputs to further enhance extreme flood forecasting.

Our development and application of DRUM open new frontiers in hydrology. Recent studies (Feng et al., 2022, 2024) have shown that integrating physical processes into DL models can improve interpretability without significantly compromising predictive accuracy. Future work could further incorporate physical constraints into the diffusion framework through physics-informed loss functions (Bastek et al., 2024; Wang et al., 2025) or guided diffusion (Chung et al., 2022; Kawar et al., 2022), thereby combining DRUM's probabilistic strengths with enhanced physical consistency. Looking ahead, we envision a new generation of hydrological models that leverage unprecedented water surface observations from the Surface Water and Ocean Topography satellite (Archer et al., 2025) and AI-powered weather prediction (Bi et al., 2023; Price et al., 2025). This convergence of technologies, particularly generative models capable of capturing both physical processes and uncertainties, paves the way for high-resolution and intelligent water resources management systems. Beyond hydrology, diffusion models hold immense potential for advancing other geoscientific domains, offering a transformative framework for simulating complex Earth system processes with greater accuracy and uncertainty quantification.

## Acknowledgments

This work was supported by the National Natural Science Foundation of China (Grant No. 42325702) and the National Key R&D Program of China (Grant Nos. 2023YFC3007700 and 2023YFC3007705). Additional support was provided by the High-level University Special Fund (Grant No. G030290001).

**Open Research**

All datasets utilized in this study are publicly available. The CAMELS dataset is publicly available from Newman et al. (2015) and Addor et al. (2017), which can be accessed at https://dx.doi.org/10.5065/D6MW2F4D. The ECMWF-IFS reforecast dataset is presented in Vitart et al. (2017) and are available at https://apps.ecmwf.int/datasets/data/s2s-reforecasts-instantaneous-accum-ecmf. ERA5-Land reanalysis data is presented in Muñoz Sabater (2019) and are available at https://doi.org/10.24381/cds.e2161bac. The code utilized in this study is freely available at https://github.com/ozg2021/DRUM.git.

## References From the Supporting Information